\documentclass[iop]{emulateapj}

\usepackage{textcomp}
\usepackage{amsmath,amssymb,amsfonts}
\usepackage{hyperref}
\usepackage{txfonts}

\usepackage{sidecap}
\usepackage{natbib}

\usepackage{chemformula} 
\usepackage{booktabs} 
\usepackage[normalem]{ulem} 
\usepackage{xcolor} 
\usepackage{colortbl} 
\usepackage[FIGTOPCAP,Large]{subfigure}

\usepackage{aas_macros} 

\newcommand*{\logten}{\mathop{\log_{10}}}

\begin{document}

\title{\bf{New versus past silica crush curve experiments:\\ application to Dimorphos benchmarking impact simulations}}

\author{Uri Malamud{$^1$}, Christoph M. Schäfer{$^2$}, Irina Luciana San Sebasti\'an{$^{3,4}$}, Maximilian Timpe{$^5$}, Karl Alexander Essink{$^5$}, Christopher Kreuzig{$^5$}, Gerwin Meier{$^5$}, Jürgen Blum{$^5$}, Hagai B. Perets{$^{1,6}$}, Christoph Burger{$^2$}}

\affil{{$^1$}Department of Physics, Technion Israel Institute of Technology, Israel}
\affil{{$^2$}Institut für Astronomie und Astrophysik, Eberhard Karls Universität Tübingen, Germany}
\affil{{$^3$}Instituto de Astrofísica de La Plata, CCT La Plata-CONICET-UNLP, Paseo del Bosque S/N (1900), La Plata, Argentina.}
\affil{{$^4$}Facultad de Ciencias Astronómicas y Geofísicas, Universidad Nacional de La Plata, Argentina}
\affil{{$^5$}Institut f\"ur Geophysik und Extraterrestrische Physik, Technische Universit\"at Braunschweig, Germany}
\affil{{$^6$}Department of Natural Sciences, The Open University of Israel, 1 University Road, PO Box 808, Raanana 4353701, Israel}

\email{urimala@physics.technion.ac.il}
	
\newpage



\begin{abstract}
Crush curves are of fundamental importance to numerical modeling of small and porous astrophysical bodies. The empirical literature often measures them for silica grains, and different studies have used various methods, sizes, textures, and pressure conditions. Here we review past studies and supplement further experiments in order to develop a full and overarching understanding of the silica crush curve behavior. We suggest a new power-law function that can be used in impact simulations of analog materials similar to micro-granular silica. We perform a benchmarking study to compare this new crush curve to the parametric quadratic crush curve often used in other studies, based on the study case of the DART impact onto the asteroid Dimorphos. We find that the typical quadratic crush curve parameters do not closely follow the silica crushing experiments, and as a consequence they under (over) estimate compression close (far) from the impact site. The new crush curve presented here, applicable to pressures between a few hundred Pa and up to 1.1 GPa, might therefore be more precise. Additionally, it is not calibrated by case-specific parameters, and can be used universally for comet- or asteroid-like bodies, given an assumed composition similar to micro-granular silica.
\end{abstract}
	



\section{Introduction}\label{S:Intro}
Small bodies in the solar system are thought to be highly porous objects, as inferred from bulk density and composition measurements of various solar system objects, including comets \citep{GroussinEtAl-2019}, asteroids \citep{BaerEtAl-2011} and Kuiper-belt objects \citep{MalamudPrialnik-2015}. Further evidence comes from direct measurements of porosity from multiple classes of meteorites \citep{Macke-2010}. Taken together, we expect nearly all small and intermediate-sized planetesimals to contain non-negligible porosity. Understanding the evolution of such bodies, and the effects of collisions, is sensitively dependent on our understanding of their porosity. In particular, this knowledge is important for planetary defense purposes. 

Note that for the remainder of this work, we will interchangeably make use of the related terms porosity ($\psi$), the volume filling factor ($\phi=1-\psi$), and distention ($\alpha=1/\phi$). While porosity is the more general term, volume filling factor is used often in laboratory work, and distention is used by impact modelers.

In the last few decades, the porosity of small planetesimals has attracted considerable attention, through theoretical studies and laboratory investigations. Compaction experiments of porous granular media play an important role in modeling collisions of small bodies, as well as in other applications like thermo-physical modeling.

When compaction is applied to computer collision simulations, a part of the kinetic energy is absorbed by compressing or deforming the porous structure, rather than contributing solely to the excavation of material. This leads to less ejecta, as well as lower ejecta velocities \citep{JutziEtAl-2008}. Additionally, a porous medium has fundamentally different material properties that considerably alter the outcomes of collisions, such as strength \citep{HaackEtAl-2020,TatsuumaEtAl-2023} or sound speed \citep{MeisnerEtAl-2012}.

Compaction in a porous medium is actuated through the change in porosity as a function of pressure. This function is sometimes referred to as the porous material equation of state, and in the impact literature, it is also more commonly referred to as its crush curve. 

Benchmarking impact simulation studies show that for a given setup and a given starting porosity, keeping all other parameters equal, the choice of the crush curve has a decisive influence on the simulation outcomes \citep{LutherEtAl-2019,LutherEtAl-2022}. Because the crush curve determines how much energy is needed for pore crushing and deformation, it changes the properties of the ejecta, the potential morphology of craters, and the overall momentum enhancement factor of the impact.

Given these arguments, the purpose of this study is to review (Section \ref{S:Past}) as well as supplement (Section \ref{S:Current}) the empirical literature on the crush curves of porous refractory materials. However, here we investigate specifically the crush curve of granular media made of various-sized silica (SiO$_2$) grains. This material has very frequently been used in experiments. Having a simple and uniform chemical makeup enables a comparison among other properties of the silica samples (size distribution / texture / method) between various studies (Section \ref{S:Comparison}). Additionally, the non-porous attributes of silica are also well known and frequently used in impact simulations of larger, consolidated objects, making it a natural and self-consistent material choice across a broad range of scales. In contrast, experiments with powders that were produced from a myriad of rocky minerals of varying complexity, origin, and properties (e.g., see \cite{HausenEtAl-2018}), do not permit this level of analysis, because the effect of composition is not decoupled from other effects. While the ultimate goal is to establish a vast catalog of crush curves for various materials, the scope of such a work is far beyond what can be achieved in a single study. Our current goal is more modest, hereafter focusing on silica only, without showing other laboratory work for other materials.

To complete the investigation, we perform our own set of impact simulations (Section \ref{S:Benchmark}), utilizing a crush curve for micro-granular silica which has never been used in the collision literature prior to this study. We advocate its use since it has a superior fit to experiments and has a wider range of applicability. We discuss the implications of our results (Section \ref{S:Discussion}), reaffirming the important role that crush curves play in collisions. Our simulations are also designed particularly for the investigation of the recent impact on the asteroid Dimorphos, which has important and topical implications for planetary defense purposes, protecting against asteroids that might potentially collide with the Earth \citep{ChengEtAl-2018}. A summary of our conclusions is finally presented (Section \ref{S:Conclusions}).

\section{Past experiments}\label{S:Past}
In order to provide the relevant context, we assemble a compilation of past experiments involving silica, which we broadly divide into two categories: experiments utilizing static compression (using a hydraulic press) and experiments utilizing dynamic compression (via high-speed collisions). With one exception, the various studies are mentioned in ascending order according to publication year.

\subsection{Static compression}
We start with the study by \cite{GuttlerEtAl-2009}, who performed both omnidirectional and unconfined unidirectional static compression. Only the results for omnidirectional compression are used in this work because they represent the situation in impacts closer than those experiments in which the material can freely creep sideways. \cite{GuttlerEtAl-2009} used monodisperse spherical dust particles with a diameter of 1.5 $\mu$m and a material density of 2000 kg $\times$ m$^{-3}$ (manufacturer: \textit{micromod Partikeltechnologie}).  

\cite{YasuiArakawa-2009} performed experiments using monodisperse 1~$\mathrm{\mu m}$ spherical silica particles mixed with water ice in different ratios and with different temperatures and compression speeds. For this work, only their results for pure silica were used with a material density of $2200~\mathrm{kg~m^{-3}}$. The pressure range investigated in their pure silica experiment was 0.2-30 MPa.

\cite{MeisnerEtAl-2012} used polydisperse angular silica particles between 0.1~$\mathrm{\mu m}$ and 10~$\mathrm{\mu m}$ with a material density of $2600~\mathrm{kg~m^{-3}}$ (manufacturer: \textit{Sigma Aldrich}), investigating the pressure range 0.25-55 kPa.

\cite{HausenEtAl-2018} measured the crush curves of various materials, but only two of their crush curves consist of pure silica. Lane Mountain (LM) sand consists of close to 100\% quartz, and therefore we include the two LM crush curves in this study (manufacturer: \textit{Lane Mountain Company}). Unlike the previous powders, the LM grain sizes are sub-mm, rather than micron-sized. Their experiments were performed by applying uniaxial compression on a servo-hydraulic test machine. The test fixture was a stainless steel cylinder, with a cylindrical pocket measuring that held the sample material. A stainless steel plunger, slightly smaller in diameter than the pocket, was pushed down into the pocket, compacting the material until the maximum capability of the machine was reached (150 MPa). The applied force and displacement of the plunger were measured with a precision load cell and extensometer, and used in order to calculate the density of the material versus the applied stress. \cite{HausenEtAl-2018} mention two LM samples, LM20/30 and LM70, however without specifying the differences between them. However, this information is available online with the manufacturer. The LM20/30 grains range in size between 0.3 and 1.18 mm, with most grains falling inside the 0.425-0.6 mm size bin. The LM70 grains range in size between 0.075 and 0.425 mm, with most grains falling in the 0.15-0.3 mm size bin. The grains are pre-tapped before the experiments in order to first obtain a bulk porosity of 43.4\% at their maximum packing density. The grain specific density is $2650~\mathrm{kg~m^{-3}}$.

The most recent fitting to the \cite{HausenEtAl-2018} data is unpublished, and was obtained via private communication with Sabina Raducan, based on recently presented work \citep{RaducanEtAl-2024c}. The full parametric fit function is of the form 

\begin{align}\label{eq:parametric}
    \alpha (P) = \left\{
    \begin{array}{ll}
        (\alpha_\mathrm{e} - \alpha_\mathrm{t}) \left(\frac{P_\mathrm{t} - P}{P_\mathrm{t} - P_\mathrm{e}}\right)^{n_1} + (\alpha_\mathrm{t} - 1) \left(\frac{P_\mathrm{s} - P}{P_\mathrm{s} - P_\mathrm{e}}\right)^{n_2} & P_\mathrm{e}<P<P_\mathrm{t}\\
        1 + (\alpha_\mathrm{t} - 1) \left(\frac{P_\mathrm{s} - P}{P_\mathrm{s} - P_\mathrm{e}}\right)^{n_2}& P_\mathrm{t}<P<P_\mathrm{s}\\
        1 & P_\mathrm{s}<P
    \end{array}
    \right.
\end{align}

where $P_\mathrm{e}$, $P_\mathrm{t}$ and $P_\mathrm{s}$ mark the elastic, transition and solid regimes, respectively, with corresponding $\alpha_\mathrm{e}$ and $\alpha_\mathrm{t}$. The precise values, along with $n_1$ and $n_2$ are given below in Table \ref{tab:Past_static}. It is important to note that the fitting was done based on the available empirical data, which exists (for the \cite{HausenEtAl-2018} experiments) up to a pressure of 150 MPa, roughly the value of $P_\mathrm{t}$ or slightly larger, but not up to $P_\mathrm{s}$. At these higher pressures, the fit was extrapolated based on other experiments made for Lunar regolith \citep{AhrensCole-1974}. Given this extrapolation, we only considered hereafter the fit values up to 150 Mpa.

Another study investigates a similar sample of quartz sand called 'Ottawa sand' (manufacturer not listed explicitly), with grain sizes between 0.425 and 0.85 mm (very similar to LM20/30) and pressures up to 345 MPa \citep{HagertyEtAl-1993}. Because this study is much older and does not provide a specific fit function in the publication, we decided to fit it ourselves in order to validate the much newer results of \cite{HausenEtAl-2018}, or look for subtle differences as a function of the grain size distribution. For our purposes, an approximate fit is sufficient, and we therefore fit this curve by entering the data in a fitting software, and selecting what we deem to be the closest fit to the data (by eye). We choose the curve provided in their figure 9, with the loose specimen, because this curve specifically has the most similar initial porosity (41.5\%) to the pre-tapped specimens of the LM sand. After exploring several options, we find a logistic mathematical function to be the most suitable for describing the data.


Finally, \cite{MalamudEtAl-2022} undertook a different approach, performing their compression experiments on hierarchical samples of pebble aggregates, rather than using homogeneous samples, as in all previous experiments. Pebbles are believed to have formed in the early solar system via extremely low-velocity collisions of micron-sized grains and might be the basic building blocks of many primitive small bodies \citep{ZsomEtAl-2010}. Each pebble is a porous aggregate of constituent micron-sized grains. Laboratory experiments show that the typical pebble intra-porosity is around 60\% \citep{WeidlingEtAl-2012}. A pile of pebbles, however, also consists of macro-porosity (i.e. large voids between the pebbles). For a loosely bound pile of pebbles, the macro-porosity is around 40\%. The pressure of self-gravity starts to become more significant in km-sized objects and can begin to eliminate macro-porosity with increasing size. Hence, the overall effective porosity of a pebble pile is extremely high, initially reaching past 75\%, but for objects that are tens of km in radius, the pressure of self-gravity is enough to eliminate all macro-porosity. In turn, objects larger than this threshold will consist of a homogeneous, rather than hierarchical pebble structure. \cite{MalamudEtAl-2022} sought to conduct compression experiments in the pressure regime that transitions from the pebble-pile state to the homogeneous state. They found this pressure range to be roughly between 3 and 250 kPa. The samples used in the experiments contained pebbles which were produced from various powders, one of which was a powder of polydisperse angular SiO$_2$ dust, with grains in the diameter range between 0.2 $\mu$m and 4.7 $\mu$m and a density of $2600~\mathrm{kg~m^{-3}}$ (manufacturer: \textit{Sigma Aldrich}). The lower size limit is determined by the mean particle size minus one standard deviation measured by counting particles, whereas the upper limit is derived by taking the mean particle size plus one standard deviation measured by volume \citep{Kreuzig2024}. 

A summary of the above experiments is assembled in Table \ref{tab:Past_static}, where the resulting expression for the volume filling factor $\phi$ is provided.

\begin{table*}
    \caption{Summary of past static compression experiments.}
    \centering
    	\begin{tabular}{|l|l|l|l|l|}
    		\hline
    		{\bf Study} & {\bf Pressure range} & {\bf Grain diameter} & {\bf Grain texture} & {\bf Crush curve analytical fit (Eq. \ref{eq:parametric} for rows 3-4)}\\ 
    		\hline
    		\cite{GuttlerEtAl-2009} & 0.3-1000 kPa  & Monodisperse, 1.5 $\mu$m  & spherical & $\phi = 0.58 - 0.46/\left(1+{\rm exp}((\logten(P)+1.886)/0.58)\right)$\\
    		\cite{YasuiArakawa-2009} & 0.2-30 MPa & Monodisperse, 1 $\mu$m  & spherical & $\phi = 1- 0.53 P^{-0.11}$\\
    		\cite{MeisnerEtAl-2012} & 2.5-55 kPa & Polydisperse, 0.1-5 $\mu$m & angular & $P_\mathrm{e/t/s}=0.01/80/600$, $\alpha_\mathrm{e/t}=1.76/1.55$, $n_{1/2}=3/4$\\
            \cite{HausenEtAl-2018} 20/30& $<$150 MPa & Polydisperse, 0.3-1.18 mm & angular & $P_\mathrm{e/t/s}=0.01/150/1400$, $\alpha_\mathrm{e/t}=1.76/1.55$, $n_{1/2}=3/4$\\
            \cite{HausenEtAl-2018} 70& $<$150 MPa & Polydisperse, 0.075-0.425 mm & angular & $\phi = 0.235384 \left(1+3386.845 P\right)^{0.115}$\\
            \cite{HagertyEtAl-1993} & 1-345 MPa & Polydisperse, 0.425-0.85 mm & angular & $\phi = 0.8247/\left(1+ {\rm exp}(-0.01126(P+78.11))\right) $\\
    		\cite{MalamudEtAl-2022} & 3.3-250 kPa & Polydisperse, 0.2-4.7 $\mu$m & angular & $\phi = 0.084 \logten(P)+0.44$\\
    		\hline
    	\end{tabular}
    	\label{tab:Past_static}
    	\newline $\phi$ is the volume filling factor and the pressure $P$ is in units of MPa. The \cite{GuttlerEtAl-2009} fit refers to omnidirectional compression.
\end{table*}

\subsection{Dynamic compression}\label{SS:Dynamic}
\cite{BeitzEtAl-2013} produced highly compressed samples of chondritic analog material \citep{MackeEtAl-2011} by high-speed impact experiments, achieving impact pressures between 90 MPa and 2400 MPa and resulting in volume filling factors from 0.7 to 0.99. The study of \cite{BeitzEtAl-2013} is slightly different from all the other studies we review in this paper. It was designed to distinguish between ordinary chondritic material (with a non-negligible amount of chondrules, which are spherules that are much larger than micron-sized dust grains) and carbonaceous chondritic material with a small or even negligible volume fraction occupied by chondrules. Hence, the samples they used are bi-modal and contain both small micron-sized silica dust grains and large mm-sized chondrule analogs made from various materials. Although we included these results here, we suggest exercising caution in the interpretation of the crush curve by \cite{BeitzEtAl-2013} due to the bi-modality of their samples, use of various materials and relatively few data points. For the micro-granular silica portion of their samples, \cite{BeitzEtAl-2013} used polydisperse angular dust, with grains in the diameter range between 0.1 $\mu$m and 5 $\mu$m and a specific density of $2600~\mathrm{kg~m^{-3}}$ (manufacturer: \textit{Sigma Aldrich}), and calculated the impact pressure according to \cite{Melosh-1989}.

\cite{SanSebastianEtAl-2020} performed dynamic crush curve experiments, following \cite{BeitzEtAl-2013}, however using a pure silica powder with grains in the diameter range between 0.1 $\mu$m and 5 $\mu$m and a density of $2600~\mathrm{kg~m^{-3}}$ (manufacturer: \textit{Sigma Aldrich}). In contrast to \cite{BeitzEtAl-2013}, the setup of the experiments changed, and the only measures taken in the experiments were the velocity of the projectile and the volume filling factor achieved after impacting the sample. Therefore, due to the lack of measurements about the volume filling factor of the pre-compressed samples and the energy dissipation of the impact, we can roughly approximate the mean impact pressure as $P_{\rm mean}= E_{\rm kin}/V_{\rm target}$ where $E_{\rm kin}$ is the kinetic energy using the projectiles velocity and $V_{\rm target}$ is the volume of the target after impact.

A summary of the dynamic compression experiments is presented in Table \ref{tab:Past_dynamic}.

\begin{table*}
    \caption{Summary of past dynamic compression experiments.}
    \centering
    	\begin{tabular}{|l|l|l|l|l|}
    		\hline
    		{\bf Study} & {\bf Pressure range} & {\bf Grain diameter} & {\bf Grain texture} & {\bf Crush curve analytical fit}\\ 
    		\hline
    		\cite{BeitzEtAl-2013} & $\sim$ 100-1000 MPa  & Polydisperse, 0.1-5 $\mu$m + mm-sized spherules & angular & $\phi = 0.568 P^{0.082}$\\
    		\cite{SanSebastianEtAl-2020} & $\sim$ 10-1000 MPa  & Polydisperse, 0.1-5 $\mu$m  & angular & $\phi = 0.598 P^{0.034}$\\
    		\hline
    	\end{tabular}
    	\label{tab:Past_dynamic}
    	\newline $\phi$ is the volume filling factor and the pressure $P$ is in units of MPa. In \cite{BeitzEtAl-2013} the volume fraction of micron-sized silica dust generally exceeded 75$\%$.  
\end{table*}

\section{Current experiments}\label{S:Current}
Given the past laboratory work reviewed in the previous section, we identify the need for new static experiments at high pressures, at least comparable to those of the past dynamic experiments. Additionally, past static experiments on micro-granular samples were performed at much lower pressures compared to mm-sized sand experiments, and never with polydispered spherical grains. These considerations motivated our present laboratory work. The new experiments performed for this study involve only static omnidirectional compression, which represents the situation in hypervelocity impacts better than unconfined static unidrectional compression where the material can creep sideways \citep[see figure 3 in][]{BlumSchraepler-2004}. As outlined below, several experiments were carried out, using different silica powders of either spherical or angular grains, as well as various grain size distributions of micro-granular powders.

\begin{figure}
    \centering
    \includegraphics[width=\linewidth]{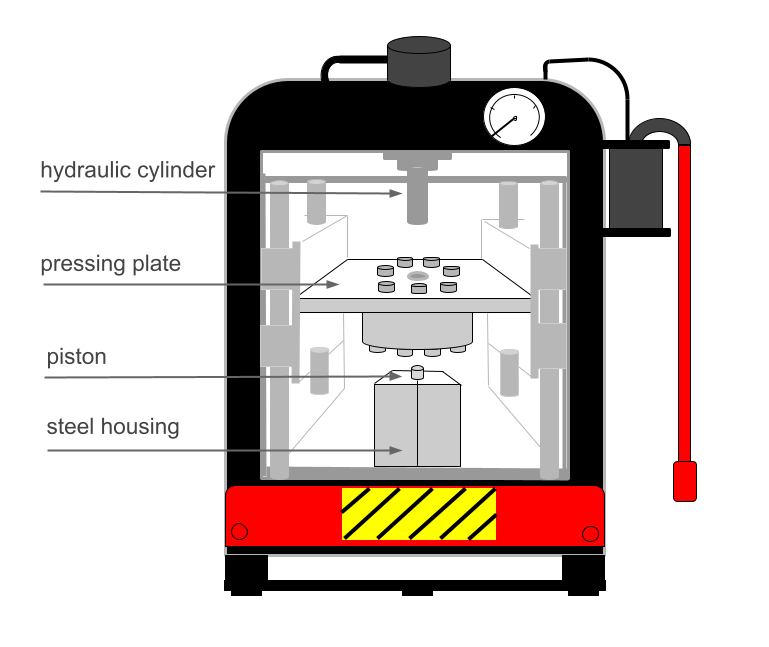}
    \caption{Schematic sketch of the experimental setup for dust compaction.}
    \label{fig:hydraulicpress}
\end{figure}

In Figure \ref{fig:hydraulicpress}, we present a schematic sketch of the hydraulic press used for the static compression of micro-granular silica-dust samples. Although the force applied on the sample is unidirectional, the corresponding compressional stress can be regarded as omnidirectional, because the sample height never exceeded its diameter so that Janssen effect \citep{Sperl2006} is minimal and the force chains become isotropic throughout the sample. With this setup, we were able to measure the pressure applied to the sample along with the change in its volume filling factor during an experimental run. The different silica powders, ranging in mass between 0.5 g and 1.3 g, were poured into a hollow steel cylinder with a diameter of 12.5 mm, between two carbide pistons. The steel housing was removable so that we could extract the sample after the compression. The volume filling factor of the sample was calculated by continuously measuring the position of the piston during compression. 

\begin{table*}
    \caption{Summary of new static compression experiments performed in this study.}
    \centering
    	\begin{tabular}{|l|l|l|l|l|}
    		\hline
    		{\bf Study} & {\bf Pressure range} & {\bf Grain diameter$^*$} & {\bf Grain texture} & {\bf Crush curve analytical fit (Eq. \ref{eq:fitfunction} for rows 1-4)}\\ 
    		\hline
    		C1 spherical powder & 2.1-1130 MPa  & 0.02-0.28 $\mu$m & spherical & $\phi_\mathrm{max} = 0.90$, $P_0 = 2.3$, $\epsilon = 4.2\times10^{-3}$, $\eta = 8.0\times10^{-4}$, $\gamma = 5.8$\\
    		C2 spherical powder & 2.1-1130 MPa & 0.1-1.06 $\mu$m  & spherical & $\phi_\mathrm{max} = 0.84$, $P_0 = 3.6\times10^{-2}$, $\epsilon = 2.5\times10^{-5}$, $\eta = 9.5\times10^{-4}$, $\gamma = 10.2$\\
            C4 spherical powder & 2.1-1130 MPa  & 0.01-2.33 $\mu$m & spherical & $\phi_\mathrm{max} = 0.88$, $P_0 = 1.1\times10^{-2}$, $\epsilon = 4.9\times10^{-6}$, $\eta = 6.5\times10^{-4}$, $\gamma = 9.0$\\
    		C6 spherical powder & 2.1-1130 MPa & 0.08-9.13 $\mu$m & spherical & $\phi_\mathrm{max} = 0.85$, $P_0 = 2.1\times10^{-3}$, $\epsilon = 1.9\times10^{-7}$, $\eta = 1.1\times10^{-3}$, $\gamma = 15.9$\\
    		Angular powder & 1.4-1060 MPa & 0.2-4.6 $\mu$m & angular & $\phi = 0.472 ~P^{0.06262}$\\
    		\hline
    	\end{tabular}
    	\label{tab:Current}
    	\newline $^*$: spherical beads are polydisperese, with the lower value of grain diameters determined by median count minus one standard deviation, the upper value of grain diameters determined by median volume plus one standard deviation; $\phi$ is the volume filling factor and the pressure $P$ is in units of MPa.
\end{table*}

\subsection{Spherical beads}

\begin{figure}
    \centering
    \includegraphics[width=\linewidth]{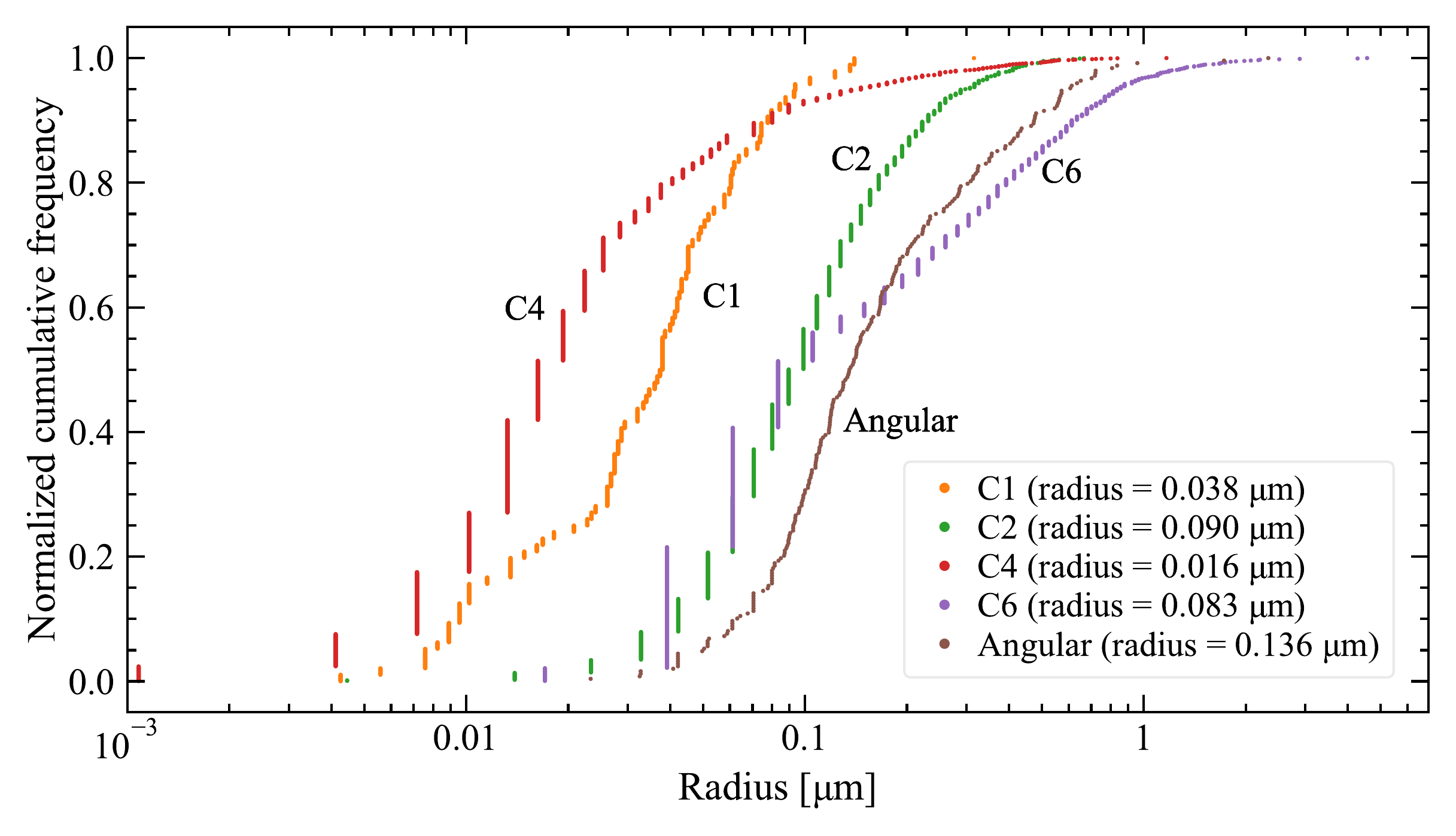}
    \caption{Cumulative size-frequency distribution of the spherical amorphous (C1-C6) and angular crystalline grains in our new experiments. The radius shown in the legend denotes the median of the count. Data were taken from \citet{Kreuzig2024}.}
    \label{fig:SphericalGrainsDistribution}
\end{figure}

We conducted compaction experiments with the hydraulic press described above in the pressure range between 2.1 MPa and 1130 MPa, using polydisperse spherical beads with a density of $2060\pm 100~\mathrm{kg~m^{-3}}$ (manufacturer: \textcopyright$\,$ \textit{Admatechs}). The spherical beads are manufactured via nucleation, sintering, and then rapid cooling, which results in an amorphous atomic structure, hence the different density compared to the angular grains, which possess a crystalline structure. The spherical powders, labeled C1, C2, C4, and C6 by the manufacturer, feature very different grain size distributions. For details see Table \ref{tab:Current} and Figure \ref{fig:SphericalGrainsDistribution}. The sample indices reflect (in increasing order) the maximal grains size in the distribution, but do not reflect the distribution width (see Section \ref{S:Comparison} for further discussion). 

\begin{figure*}
    \centering
    \includegraphics[width=0.49\textwidth]{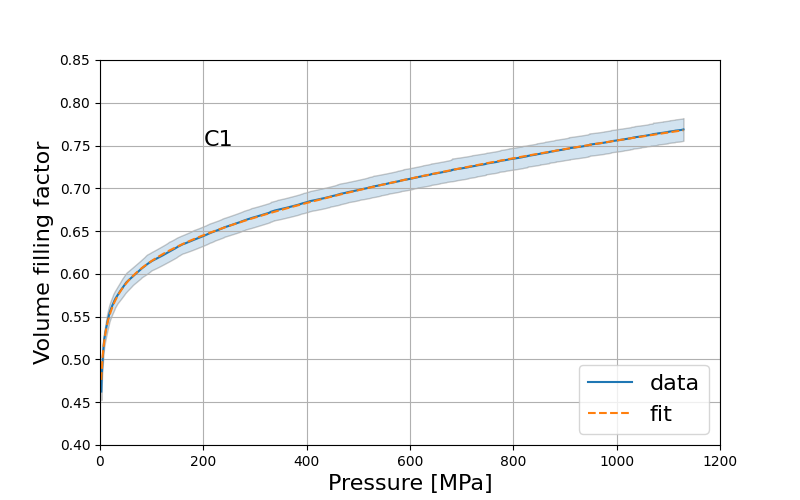}
	\includegraphics[width=0.49\textwidth]{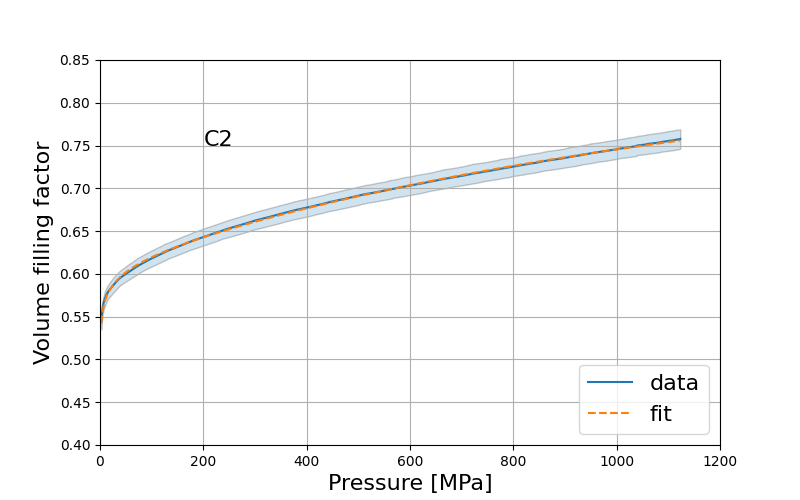}
	\includegraphics[width=0.49\textwidth]{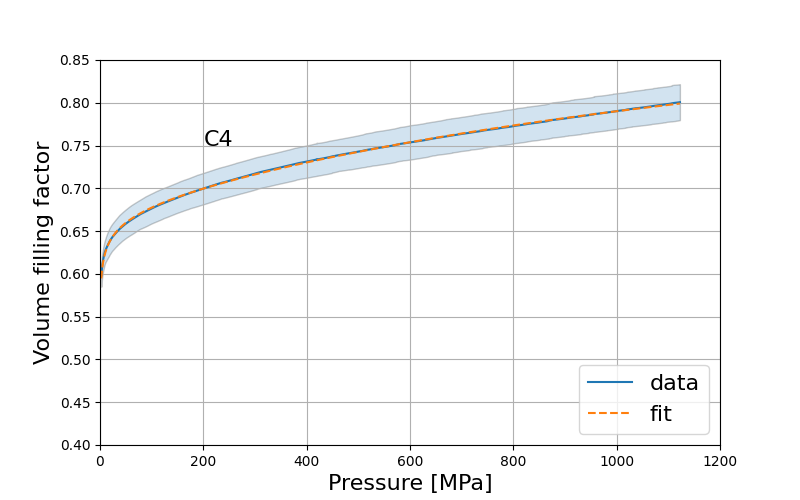}
	\includegraphics[width=0.49\textwidth]{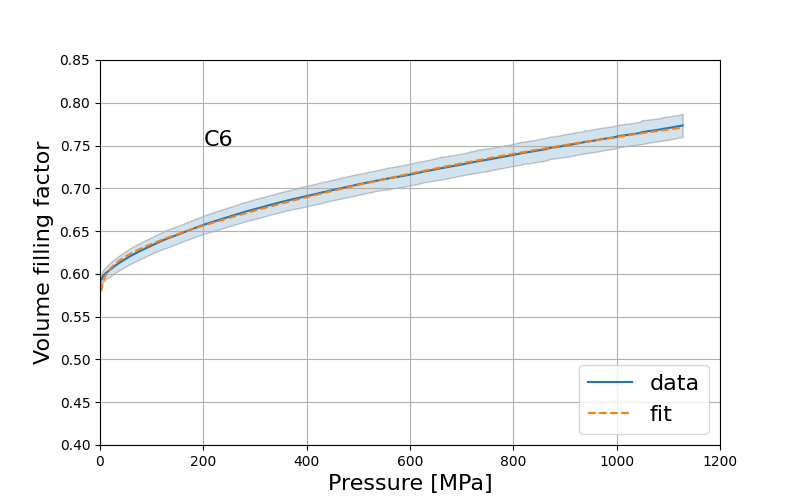}
    \caption{Compression curves and fits of Eq. \ref{eq:fitfunction} of the spherical silica samples C1 (upper-left), C2 (upper-right), C4 (lower-left) and C6 (lower-right). The solid blue curves denote the average crush curves of the five experimental runs. Shaded regions represent the standard error of the mean values. The orange dashed curves show the best fits according to Eq. \ref{eq:fitfunction}}.
    \label{fig:app2}
\end{figure*}

The sample of known mass was introduced into the sample holder whereupon the continuously increased force was applied and the thickness of the sample was simultaneously measured with a calibrated camera setup. For each of the four samples, we performed five individual compression experiments whose results are shown in Figure \ref{fig:app2}. The solid blue curves denote the mean volume filling factor (averaged over the five measurement runs) as a function of applied pressure, whereas the blue-shaded regions represent the standard error of the mean values.


In order to obtain an analytical fit for the crush curves shown in Figure \ref{fig:app2}, we first attempted to use the basic functional relation between the volume filling factor $\phi$ and the omnidirectional pressure $P$ from \cite{TatsuumaEtAl-2023}, which reads
\begin{equation}
    \frac{1}{\phi}-\frac{1}{\phi_\mathrm{max}} = \left(\frac{\delta}{P}\right)^{1/3},
    \label{eq:tatsuuma}
\end{equation}
with $\phi_\mathrm{max}$ and $\delta$ being the maximum volume filling factor achievable and a constant that depends on the grain size, the surface energy, and the critical rolling displacement \citep[see][for details]{TatsuumaEtAl-2023}. However, we found a good agreement either only for small or only for high pressures. Thus, we modified Eq. \ref{eq:tatsuuma} by expanding $\delta$ to a pressure-dependent value and by replacing the fixed exponent also by a pressure-dependent value. Introducing three new parameters $\epsilon, \eta, \gamma$, our fit function reads
\begin{equation}
     \frac{1}{\phi}-\frac{1}{\phi_\mathrm{max}} = \left(\frac{P_0 + \epsilon P}{P}\right)^{\frac{1+\eta P}{\gamma}}.
    \label{eq:fitfunction}
\end{equation}
With Eq. \ref{eq:fitfunction}, we obtained excellent fits for all four compression curves as shown in Figure \ref{fig:app2}.

\subsection{Angular particles}
We also performed compaction experiments with the hydraulic press described above (see Figure \ref{fig:hydraulicpress}) in the pressure range 1.4-1060 MPa, using micron-sized irregular silica grains (manufacturer: \textit{Sigma Aldrich}) with a diameter range between 0.2 and 4.7~$\mu$m and a grain density of $2600~\mathrm{kg~m^{-3}}$ (Table \ref{tab:Current}). The diameter of the samples was 12.5 mm with a typical mass of $\sim 1.3$ g for low-pressure experiments and ranging from 0.5 g to 1.0 g for high-pressure experiments. 

The compression experiments were performed with a previous version of the setup described above. The height of the sample could only be determined after the pressure was released. In the analysis of the compression experiments, a simple power-law relation between the volume filling factor and the applied omnidirectional pressure,
\begin{equation}
    \phi=\lambda \; P^{\nu},
    \label{eq:powerlaw}
\end{equation}
was found to fit the data best, with $\lambda=0.472$ and $\nu=0.06262$.

\section{Comparing crush curves}\label{S:Comparison}
\begin{figure*}
    \includegraphics[scale=0.78]{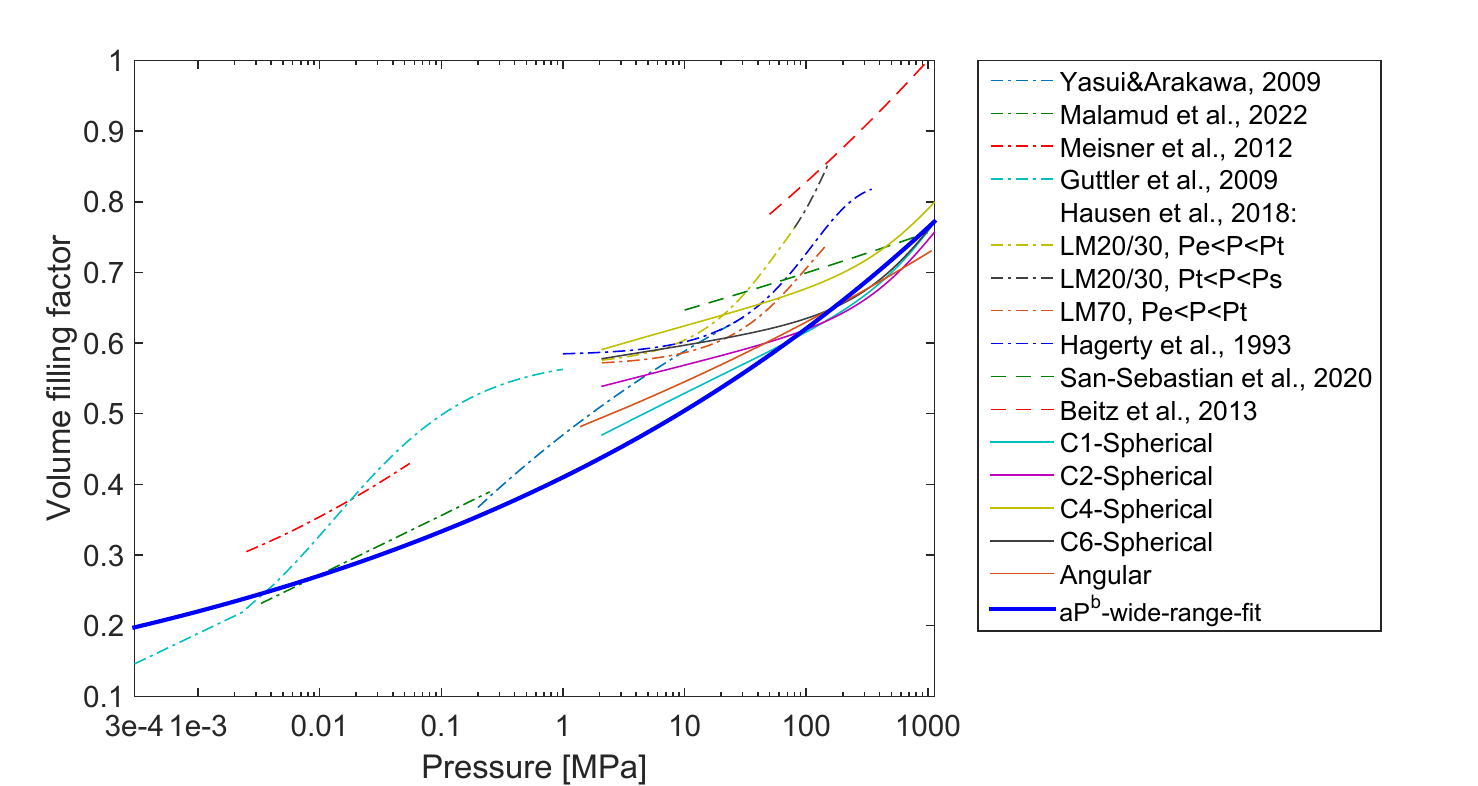}\label{fig:CurvesWithWideFit}
    	
    \caption{Silica experimental crush curves. Line style: solid (thin) -- new static experiments; dash-dotted (thin) -- past static experiments; dashed (thin) -- past dynamic experiments; solid (thick) blue -- wide range power-law fit for micro-granular silica.}
    	
    \label{fig:CrushCurves}
\end{figure*}

Figure \ref{fig:CrushCurves} shows all past (Section \ref{S:Past}) as well as new (Section \ref{S:Current}) experimental crush curves described previously for silica. The new experiments are shown by thin solid lines, whereas dash-dotted and dashed lines depict past static and dynamic experiments, respectively. Note that the static experiments cover a very wide range of pressures, hence the x-axis is shown in logarithmic scale. Although differences among the various experiments are to be expected, and several distinct mathematical functions also fit the crush curves, it is interesting that the overall trend in the data is approximately log-linear over the entire pressure range.

The dynamic crush curve from \cite{BeitzEtAl-2013} most notably deviates from the rest, displaying highly enhanced compressibility. However, as already explained in Section \ref{SS:Dynamic}, their study is unique through its use of non-pure SiO$_2$ dust samples, including large spherules, and it has been flagged as possibly inconsistent with the other studies, which indeed turned out to be the case. The second dynamic crush curve by \cite{SanSebastianEtAl-2020} is nevertheless very similar to the other static crush curves, demonstrating that the experimental details, like the initial conditions as well as grain size distribution and texture, might be more influential than the experiment methodology (static versus dynamic). We further note that if we extrapolate both the \cite{MalamudEtAl-2022} and the new angular crush curve to high/low pressures, respectively, they both represent a reasonable fit to the micro-granular experiments over the full pressure range, spanning over six orders of magnitude. The former slightly underestimates the volume-filling factor at high pressures, and the latter slightly overestimates the volume-filling factor at low pressures.

We find that there is a particularly high importance to the width of the grain size distribution. The crush curves of our new spherical powders labeled C1 to C6 exemplify this finding. All the new crush curves used in this paper were obtained with a single compression cycle and no relaxation. Within the uncertainties, the crush curves of samples C1, C2, and C6 roughly overlap for pressures above $\sim 100$ MPa. In contrast, the crush curve of sample C4 lies significantly above the other three for most of the pressure range investigated. The softness of the C4 dust is caused by contamination with ultra-small particles (smaller than those of even C1), as high-resolution electron-microscope images revealed \citep{Kreuzig2024}. The sorting of the volume filling factor at high pressures (C4$>$C6$>$C1$>$C2) seems to reflect the width of the size distribution, which decreases exactly in the same order (see Table \ref{tab:Current}). \cite{Anzivino-2023} showed that the value of the filling factor for the densest possible packing of polydisperse spheres increases with increasing width of the size distribution. Thus, at the high-pressure end of our measurements, we are observing a re-arrangement of the spheres for which the smallest particles are pushed into the voids between the large grains. This is supported by the post-compression inspection of the samples for which the SEM images showed no signs of particle break-up. Thus, pressures of $\lesssim$1 GPa are insufficient to crush individual micron-sized silica particles.

In contrast, the mm-sized LM and Ottawa sand grains investigated in past compression experiments demonstrate a much smaller porosity even at modest pressures of merely 150 MPa (see Table \ref{tab:Past_static}), indicating that some form of breakage likely does take place. Breakage of brittle materials is a size-dependent effect, with larger grains developing fractures earlier and breaking at lower pressures compared to small grains of the same material (from equation 6 and table III of \cite{BenzAsphaug-1999}, the breakage pressure differs by about 1 order of magnitude between 1 $\mu$m and 0.5 mm grains). Additionally, it is also possible that the angular shape of the sand grain further resulted in some alignment or breakage of sharp tips/edges, magnifying the breakage effects. \cite{HausenEtAl-2018} did not discuss having performed an SEM analysis on their sand samples in order to elucidate either of these possibilities.

Moreover, as the example of the C4 powder shows, it is easier to achieve a given volume filling factor if the size distribution is so wide that the inter-particle contacts are dominated by small spheres located between large ones for which a re-arrangement by rolling is much easier than for contacts between similar-sized particles. This was also shown by \cite{Kreuzig2024} who found that C4 samples possess much lower tensile strengths than expected for a narrow size range. In conclusion, a pivotal factor in the compressibility of our crush curves is the width of the size distributions, and not merely the minimum or maximum of the grain sizes (as was demonstrated for our micro-granular samples). However, the highly enhanced compression of the \cite{BeitzEtAl-2013} dynamic crush curve is likely the result of not only using an additional material besides SiO$_2$ but also of having an extremely wide, bi-modal size distribution, where small grains can get between larger ones, in addition to breakage of larger grains which contributes to the same effect, as discussed in the previous paragraph in the context of the Lane Mountain and Ottawa sand samples.

Besides the various thin lines in Figure \ref{fig:CrushCurves}, we also show a thick blue solid line, obtained using a simple analytic power-law expression:
\begin{equation}
    \phi = a P^b ,
    \label{eq:new_crush_curve}
\end{equation}
with parameters $a=0.41$ and $b=0.09$ and $P$ in MPa. This line presents a compromise that most closely follows the overall trend for the micro-granular crush curves, deviating only slightly from the \cite{MalamudEtAl-2022} curve at low and mid pressures and following closely the angular as well as spherical crush curves at high pressures. Its appropriateness over a wide pressure range and its mathematical simplicity make it an appealing choice for its use in a variety of applications. Note that when the initial volume filling factor of the sample is larger than 0.2, the fit should be replaced with a horizontal (constant volume filling factor) line up to an appropriately large pressure to enact compression, as to represent the elastic regime. This elastic regime is shown and elaborated on in Figure \ref{fig:CurvesWithQuad}, for particular assumptions on the initial porosity. In Section \ref{S:Benchmark}, we use the wide-range fit in new benchmark impact simulations. In Section \ref{S:Discussion} we discuss why it differs significantly from typical parametric quadratic crush curves that are quite often used in impact studies throughout the literature.

\section{Benchmarking impact simulations}\label{S:Benchmark}
In order to study the influence of the new micro-granular silica crush curve from Equation \ref{eq:new_crush_curve} on recent simulations in the context of the DART and Hera missions, we performed simulations based on the parameters and setup from the DART impact benchmark from \cite{LutherEtAl-2022}. This study has methodically investigated the effect of various model realizations, and also various codes, including the \texttt{miluphcuda} \citep{2016A&A...590A..19S,2020A&C....3300410S} SPH code which is used here. Although newer studies \citep{RaducanEtAl-2024a,RaducanEtAl-2024b} use more up-to-date model parameters compared to the older setup of \cite{LutherEtAl-2022}, e.g. a considerably higher density of the consolidated material, or explicitly accounting for boulders, here we are merely interested in isolating the effect of the crush curve. Hence, for our purpose, the most important aspect of the new benchmarking simulations is to keep all other parameters equal, while changing the crush curve. Based on the results presented here, more accurate future studies may then be performed, taking advantage of the new crush curve, provided that the simulated porous material is sufficiently similar to micro-granular silica.

The efficiency of momentum transfer from the impactor on the target is usually expressed in terms of the so-called $\beta$ factor. The general definition of $\beta$ is given in \cite{2021PSJ.....2..173R}. We use the simplified equation for the impact direction anti-parallel to the surface normal at the point of impact and neglect any asymmetry in the ejecta plume to compare to the values found by \cite{LutherEtAl-2022}. With these simplifications, the $\beta$-factor can be expressed in terms of the impactor momentum $p_\mathrm{imp}$, the ejecta momentum $p_\mathrm{ej}$ and the change of the momentum of the asteroid $\Delta p_\mathrm{ast}$
\begin{align}
    \beta = 1 + \frac{p_\mathrm{ej}}{p_\mathrm{imp}}= \frac{\Delta p_\mathrm{ast}}{p_\mathrm{imp}} \approx \frac{m_\mathrm{ast} \Delta v_\mathrm{ast}}{m_\mathrm{imp} v_\mathrm{imp}}.
\end{align}
Here, $m_\mathrm{imp}$ and $m_\mathrm{ast}$ denote the masses of the impactor and the asteroid, respectively, $v_\mathrm{imp}$ is the relative impact velocity, and $\Delta v_\mathrm{ast}$ is the change in the orbital velocity of the asteroid.

A 1 m in diameter, 600 kg massive spherical aluminum projectile impacts vertically (head-on) into a porous target at a speed of 6 km $\times$ s$^{-1}$. We vary the cohesion (0, 1.4, 10 and 100 kPa) and the initial porosity (42\% and 50\%, corresponding to the distention parameter $\alpha_0=1/\phi_0=1.72$ and $2$) of the target, and use a constant coefficient of friction of 0.77. These, and all other parameters were set to the same values that were used in the \cite{LutherEtAl-2022} benchmark study with \texttt{miluphcuda}. Likewise, for the yield strength we similarly apply a Lundborg model with a high-pressure limit of 1 GPa.

The SPH simulations in \cite{LutherEtAl-2022}, however, used (in commonality to most impact studies in recent years) a single power-law slope quadratic crush-curve:
\begin{align}
    \alpha (P) = 1 + (\alpha_\mathrm{0} - 1) \left(\frac{P_\mathrm{s} - P}{P_\mathrm{s} - P_\mathrm{e}}\right)^n,
    \label{eq:crush_curve_jutzi}
\end{align}
where $n=2$, $P_\mathrm{s}$ (1.3 GPa) is the coefficient for the solid pressure, $P_\mathrm{e}$ (100 Pa) is the coefficient for the elastic pressure and $\alpha_0=1/\phi_0$ is the initial distention (see \citealt{JutziEtAl-2008, 2009Icar..201..802J} for a detailed description of the P-$\alpha$~model). Note that Eq. \ref{eq:crush_curve_jutzi} is identical to the mid row in the more general parameterized fit from Eq. \ref{eq:parametric}. We use the Tillotson equation of state (EOS) for both materials, impactor and target. The parameters are listed in Table~\ref{tab:tillo_params}.
\begin{table}
    \caption{Parameters for the Tillotson EOS}
    \centering
    	\begin{tabular}{|l|l|l|}
    		\hline
    		{\bf Parameter} & {\bf Target} & {\bf Projectile} \\ 
      \hline
      Reference density (kg $\times$ m$^{-3}$) & 2700 & 1000 \\
      Bulk modulus (GPa) & 26.7 & 7.6 \\
      Tillotson B (GPa) & 26.7 &  0 \\
      Tillotson E$_0$ (J $\times$ kg$^{-1}$) & 4.87 $\times$ 10$^8$ & 0   \\
      Tillotson a & 0.5 & 0 \\
      Tillotson b & 1.5 & 0 \\
      Tillotson alpha & 5 & 0  \\
      Tillotson beta & 5 & 0  \\
      Energy of incipient vaporization (J $\times$ kg$^{-1}$) & $4.72 \times 10^6$ & $4.72 \times 10^6$   \\
      Energy of complete vaporization (J $\times$ kg$^{-1}$) &  18.2 $\times$ 10$^6$ & 18.2 $\times$ 10$^6$ \\
    \hline
   \end{tabular}
    	\label{tab:tillo_params}
\end{table}

For the new simulations, we applied the micro-granular silica crush-curve derived from the wide-range $\phi = a P^b$ fit (Equation \ref{eq:new_crush_curve}). It is important to note that compression at the lowest pressures in the experiments was only possible when the corresponding sample had an initially small volume filling factor (or large porosity, equivalently). Otherwise, in general, for every given initial porosity in the asteroid, there exists a threshold pressure below which one cannot compress the porous material at all. Therefore, in our new implementation, the crush curve has a slope of zero below this threshold value (see Figure \ref{fig:QuadCrushCurves}), in close similarity to the elastic regime in the quadratic crush curve. Unlike in \cite{LutherEtAl-2022}, here we did not consider initial porosities of 30\% or lower -- because it has already been found that the typical pressures during the impact necessarily induce negligible compression in this case, therefore we deem such investigation unnecessary. Also, rubble pile porosities are typically expected to be 40\% or more, given Dimorphos' size.

The initial particle distribution consisted of the spherical impactor, whose diameter was resolved by approximately six particles, and a flat, half-sphere target. To increase the accuracy at the impact point, the number of particles decreased radially from the impact point. The particle number density in the impact point matched the number density of the impactor to avoid numerical issues.

The simulation times varied since they were stopped as soon as a convergence of the $\ beta$ factor was achieved. The shortest simulation time was 4 s. To check for consistency and to compare to later times, we also ran some simulations up to 9 s. The default resolution was 482000 particles, however, we ran also two simulations with a higher resolution with the new crush curve that used slightly more than 10$^6$ particles.

The results of the simulations are presented in Table \ref{tab:beta} in terms of the $\beta$ factor, and compared with past results using \texttt{miluphcuda} in the \cite{LutherEtAl-2022} study.

\begin{table}
    \caption{Summary of simulation outcome - $\beta$-factor comparison for two different crush-curves}
    \centering
    	\begin{tabular}{|l|l|l|}
    		\hline
    		{\bf Parameters} & {\bf $\beta$ \citep{LutherEtAl-2022}} & {\bf $\beta$ new crush-curve} \\ 
      \hline
      $\alpha_0 =1.72$ &  &  \\
      ~~~cohesion (kPa) & &  \\
      ~~~~~0 & n/a & 3.13 \\
      ~~~~~1.4 & 2.00 & 2.06  \\
      ~~~~~10 & 1.74 & 1.74  \\
      ~~~~~100 & 1.52 &1.51  \\
      \hline
      $\alpha_0 =2 $& &  \\ 
      ~~~cohesion (kPa) & &  \\
      ~~~~~0 & n/a & 2.66\\
      ~~~~~1.4 & 1.85 & 1.74 \\
      ~~~~~10 & 1.62 & 1.52 \\
      ~~~~~100 & 1.42 & 1.36 \\
    \hline
   \end{tabular}
    	\label{tab:beta}
\end{table}

\section{Discussion}\label{S:Discussion}
As shown in the previous section, simulations using the new micro-granular silica crush curve result in $\beta$ coefficients that are very similar to those of \cite{LutherEtAl-2022} when the initial porosity is 42\%, however when it is raised to 50\%, the new $\beta$ values are notably smaller. These differences can be understood by directly comparing crush curve functions.

\begin{figure*}
    \includegraphics[scale=0.78]{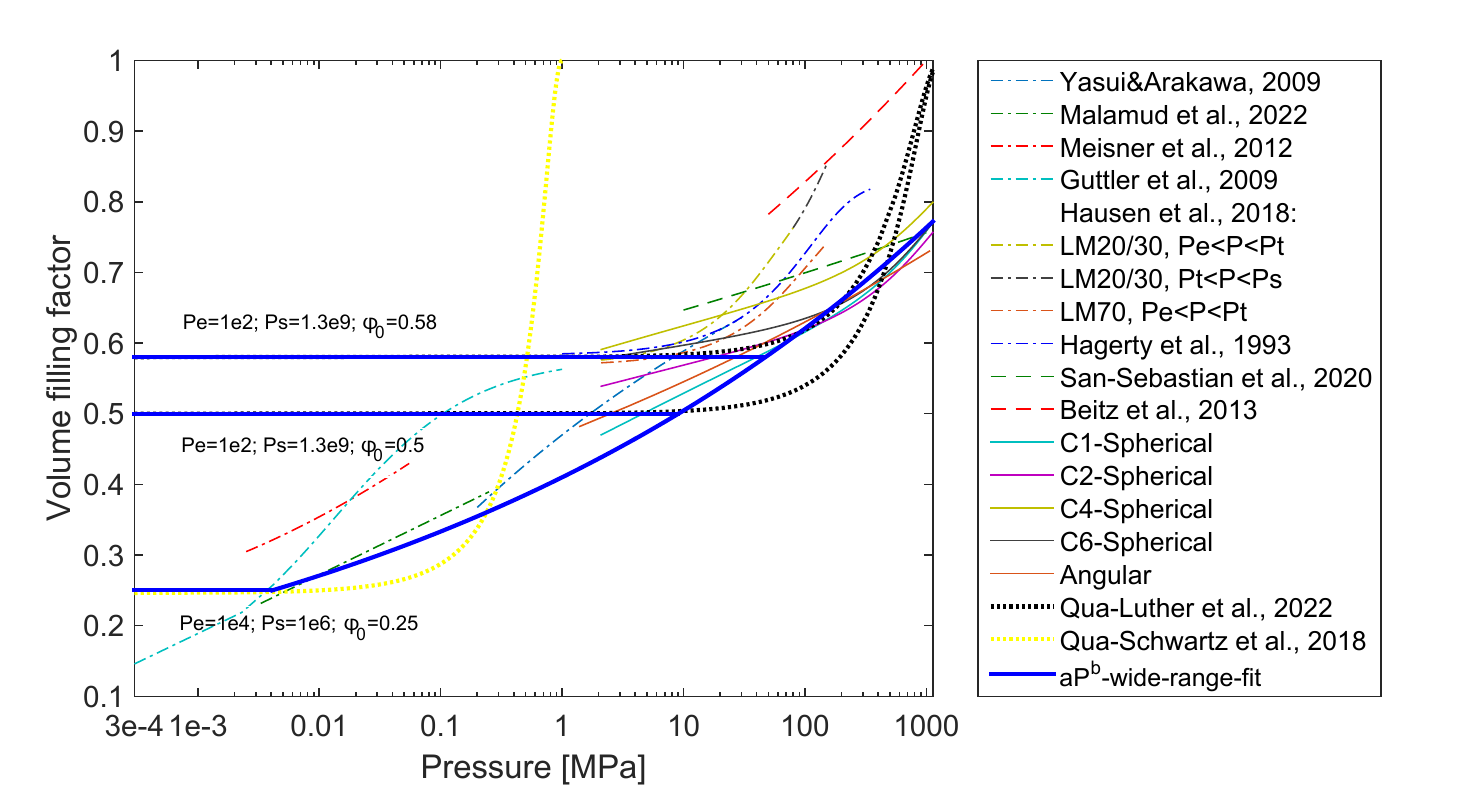}\label{fig:CurvesWithQuad}
    	
    \caption{Typical quadratic crush curves versus wide range power-law fit for micro-granular silica. Line style: same as in Figure \ref{fig:CrushCurves} but also with yellow (comet) and black (asteroid) dotted thick lines, depending on a given initial porosity.}
    	
    \label{fig:QuadCrushCurves}
\end{figure*}

Figure \ref{fig:QuadCrushCurves} is similar to Figure \ref{fig:CrushCurves}, showing the experimental crush curves denoted by the thin lines, but now also thick lines that denote some examples of parametric quadratic crush curves which are typical of those often used in impact studies in the literature. The thick black dotted lines are for asteroid-like objects, given typical parameters ($P_{\rm e}$ and $P_{\rm s}$) from \cite{LutherEtAl-2022} and the thick yellow dotted line is plotted for typical parameters of comet-like objects \citep{SchwartzEtAl-2018}. In the legend both curves are labelled via 'Qua', short for quadratic. Note that for the asteroid-like curves, two different realizations are shown, based on a different choice of the initial porosity or volume filling factor (labeled in the figure as $\phi_0$). These two choices correspond to the two initial distention values ($\alpha_0=1/\phi_0$) chosen by \cite{LutherEtAl-2022}, and are expressed in Equation \ref{eq:crush_curve_jutzi}. 

These quadratic curves for asteroids and comets do not accurately follow the experimental results for a wide range of pressures, even though they are tuned by having a different set of $P_{\rm e}$-$P_{\rm s}$ parameters in each case. The disagreement becomes larger as one deviates from the initial volume filling factor: at first, the parametric curves under-estimate the volume filling factor, and then towards larger pressures they highly over-estimate it, and even reach full compression although the empirical experiments have indicated otherwise. Additionally, Figure \ref{fig:QuadCrushCurves} shows the new crush curve tested in this study, for micro-granular silica, using Equation \ref{eq:new_crush_curve}. It is important to note again that here we do not consider different materials other than silica (e.g. pumice or Lunar regolith), which respond differently to compression. It is also important to note that larger grain sizes, like in the \cite{HagertyEtAl-1993}, \cite{BeitzEtAl-2013} or \cite{HausenEtAl-2018} silica crush curves (see Section \ref{S:Past}), affect compression. In that sense, our investigation, which only utilizes the new crush curve for micro-granular silica, is limited. However, it is easy to see that the quadratic crush curves that we have chosen as characteristic past examples for asteroid- and comet-like materials, are actually more compatible in their mid-sections with the micron-sized silica empirical data and not the mm-sized silica samples, which justifies our choice to utilize the new micro-granular silica crush curve.

As already explained in Section \ref{S:Benchmark}, compression is relevant beyond a certain pressure threshold, depending on the initial porosity. Below this threshold, there might only be some limited consolidation or a completely elastic deformation. This is why in practice, the implemented crush curves (depicted here as well by thick blue solid lines) depend on the initial porosity and are characterized by a horizontal line on the x-axis between zero and the threshold point, only after which they start to notably differ from the quadratic crush curves.

Figure \ref{fig:QuadCrushCurves} shows that the differences between the old quadratic and the new crush curves are very small when the initial porosity is set to 42\%. Only beyond a pressure of a few 10$^2$~MPa do we begin to see a notable difference, the old quadratic crush curve giving enhanced compression. In the benchmarking impact simulations, however, very few SPH particles actually experience greater pressures than these, and therefore the overall outcomes are extremely similar, manifested through similar $\beta$ values.

\begin{figure}
    \centering
    \includegraphics[width=0.48\textwidth]{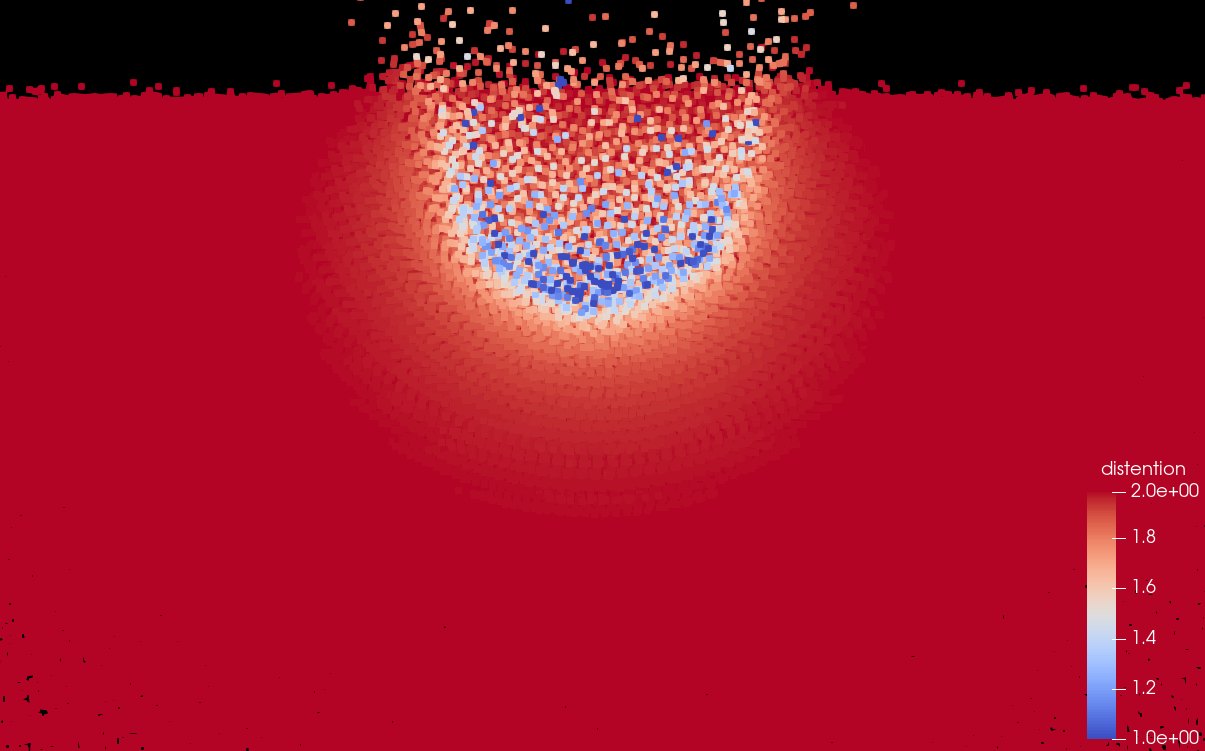}
	\includegraphics[width=0.48\textwidth]{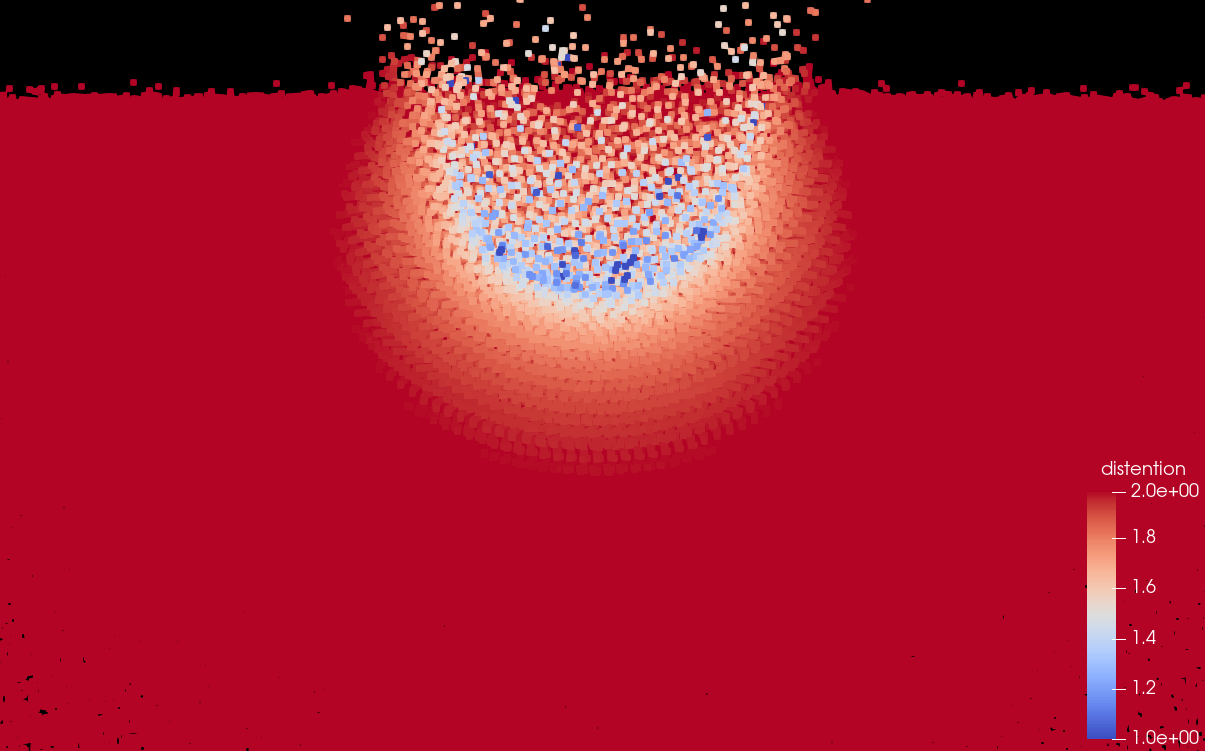}
    \caption{A side view of a clipped half hemisphere, showing the distention ($\alpha=1/\phi$) in the quadratic (top) versus new (bottom) crush curve, spanning the range between a maximum value of 2 (red - fully uncompressed with $\phi=\phi_0=0.5$) and a minimum of 1 (blue - fully compressed). Material adjacent to impact site is more compressed when using the new crush curve.}
    \label{fig:SPH_1-2}
\end{figure}

\begin{figure}
    \centering
    \includegraphics[width=0.48\textwidth]{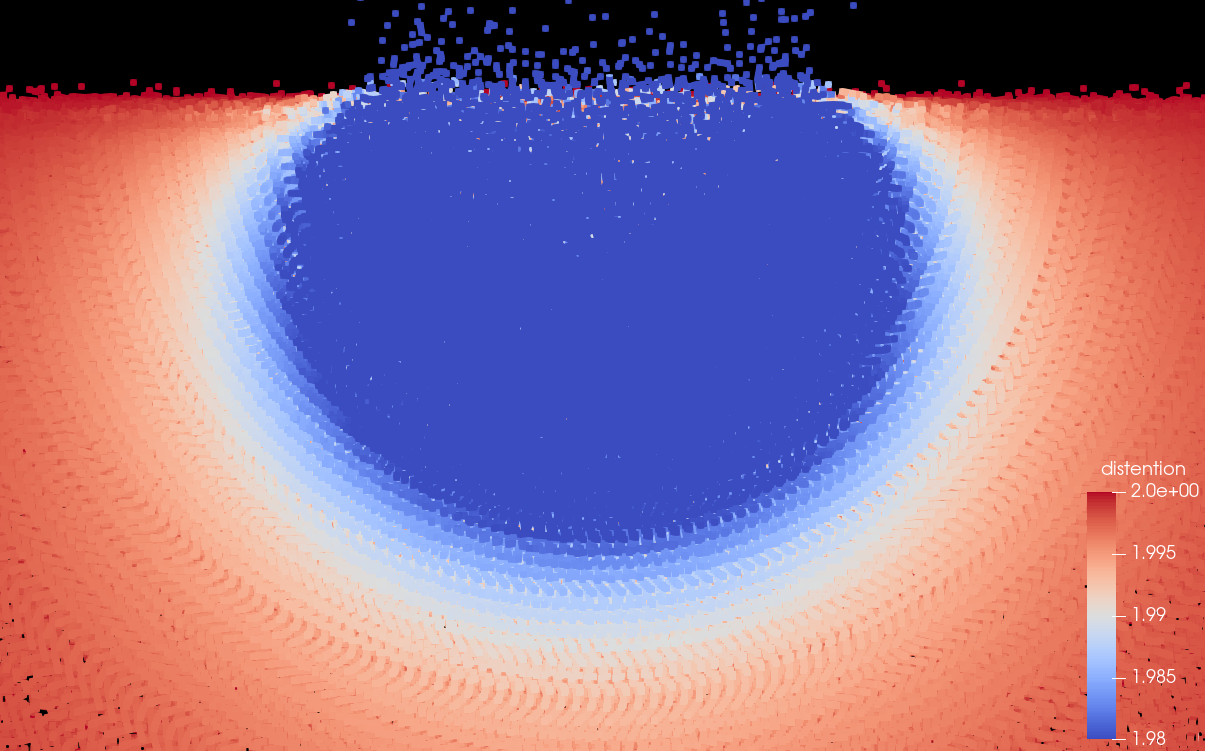}
	\includegraphics[width=0.48\textwidth]{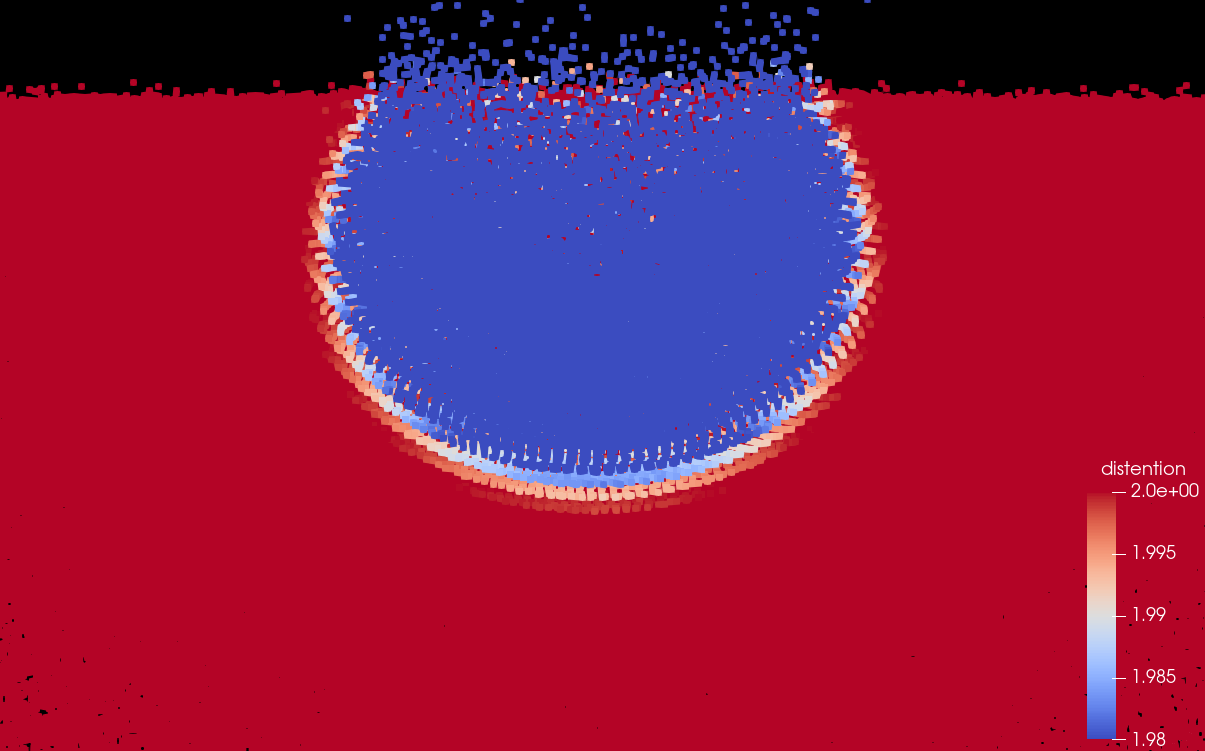}
    \caption{Same as Figure \ref{fig:SPH_1-2}, now showing a narrower distention range between a maximum value of 2 (red - fully uncompressed with $\phi=\phi_0=0.5$) and a minimum of 1.98 (blue - a 1\% decrease). A large volume far from the impact site is very slightly compressed only when using the quadratic crush curve.}
    \label{fig:SPH_1_98-2}
\end{figure}

When the initial porosity is set to 50\%, however, the simulation outcomes are quite different, the new crush curve generally resulting in lower $beta$ values. Figures \ref{fig:QuadCrushCurves}-\ref{fig:SPH_1_98-2} explain why. Since the quadratic crush curve is characterized by substantially lower volume filling factors compared to the new micro-granular silica crush curve at pressures in the range 10-500~MPa (see Figure \ref{fig:QuadCrushCurves} for $\phi_0=0.5$), material near the impact site in this pressure range exhibits considerably less compression. However, in the highest pressure range, the quadratic crush curve is again different and has a much steeper slope. Unlike the new crush curve, it can reach full consolidation.

In Figure \ref{fig:SPH_1-2} we show a snapshot from the simulation setup, corresponding to time $t=0.012$~s, featuring a side view of a clipped half hemisphere. Already at this early time after the impact, the asteroid material is subjected to the largest pressures in the simulation. The top image corresponds to the simulation with the quadratic crush curve while the bottom image to the new crush curve. The fully red pixels indicate completely uncompressed material with the 50\% initial porosity (distention $\alpha=2$), whereas the deep blue pixels represent fully consolidated material. As expected from the previous paragraph, it is clearly visible that the material surrounding the crater is considerably less compressed for the quadratic crush curve, showing a sharp boundary between the crater walls and the surrounding volume, while the bottom image features a more moderate distention gradient and a much larger volume is significantly compressed. On the other hand, the bottom image shows considerably fewer consolidated SPH particles (deep blue pixels), and those that do show belong to the kinetic projectile itself, whereas in the top image, a limited number of SPH particles that directly interacted with the projectile during the incipient contact with the surface have also been fully compressed. The immediate surroundings of the impact site are overall more significantly compressed using the new crush curve.

In contrast, as one moves further away from the impact site, the material is very slightly compressed when using the quadratic curve, whilst it remains completely uncompressed using the new crush curve. This is shown in Figure \ref{fig:SPH_1_98-2}, which is similar to Figure \ref{fig:SPH_1-2}, but here we set the lower limit of the distention (deep blue pixel) to represent merely a 1\% decrease from the initial value of 2. Using this cutoff, the slight compression is noticeable only in the upper image. It is possible to understand its origin when comparing the two crush curves for $\phi_0$ in Figure \ref{fig:QuadCrushCurves}. The quadratic curve's slope is nearly zero at low pressures, but not precisely zero, which enables some compression on the level of approximately 1/2\%. This compression is almost negligible, and the extra compression near the impact site using the new crush curve is much more significant in the overall balance. Since the compression is enhanced overall using the new crush curve, less energy goes to the ejecta. In turn, $\beta$ is expected to be smaller, which is exactly what we find with our benchmark simulations.

It should be noted that if we had compared the quadratic crush curves to characteristic mm-sized silica sands (instead of our new crush curve for micro-granular silica) the gap between the \cite{LutherEtAl-2022} results and our new results would have increased. Figure \ref{fig:QuadCrushCurves} clearly shows that the former sands are even more compressible. The simulations would have resulted in greater compression and in turn even lower $\beta$. This would have been the outcome for both assumptions of initial porosity (42\% and 50\%, and not merely 50\% as with the micro-granular crush curve). 

What could be the implications of using the new micro-granular silica crush curve for the DART impact interpretation? Didymos requires a cohesion of at least tens of Pa in order to maintain structural stability against a rapid rotation period of 2.26~h \citep{ZhangEtAl-2021}. So far, however, simulations of the Dimorphos impact suggested that reaching a $\beta$ as large as was observed \citep{ChengEtAl-2023}, requires a very low cohesion under a few Pa by the most recent estimates \citep{RaducanEtAl-2024a,RaducanEtAl-2024b}. Some speculations have been made regarding why such a disparity might exist between Didymos and Dimorphos \citep{RaducanEtAl-2024a}, however, an open mind must be kept until Hera is able to better constrain the physical properties of both asteroids. For the new crush curve we introduced here we also performed some cohesionless simulation runs. As expected, Table \ref{tab:beta} shows that $\beta$ is largest in these cases, and might be compatible with the impact observed outcomes, but more difficult to reconcile with the cohesion of Didymos.  

Previously, \cite{LutherEtAl-2022} have already robustly demonstrated that the crush curve constitutes as a key relation that affects the outcome of simulations. Our new results show that crush curves should be very carefully chosen and that our new crush curve in particular can lead to a different $\beta$, with all other parameters staying the same. \cite{LutherEtAl-2022,RaducanEtAl-2024a,RaducanEtAl-2024b} used virtually the same quadratic crush curve, with only a negligible difference between the studies. Therefore, the more recent publications that look at the longer-term effect on Dimorphos may also benefit from reviewing past choices for the crush curve.

An additional point to consider is the universality of the crush curve. While the quadratic crush curve is parametric and requires using a set of what is essentially ad-hoc parameters ($P_{\rm e}$ and $P_{\rm s}$), the new crush curve is not case-specific. It applies equally to comets, asteroids and dwarf planets, as long as the material is sufficiently similar to micro-granular silica. Of course, similar derivations may also be drawn for other materials. 

This last comment is especially true for comets or relatively small Kuiper belt objects which are highly porous. These objects are typically dominated by refractory materials (see discussion and examples in \cite{MalamudEtAl-2024}) and therefore an empirical crush curve for micro-granular silica may be reasonably judicious \citep{MalamudEtAl-2022}. With the quadratic crush curve, the parameters must first be chosen, and those from \cite{SchwartzEtAl-2018} in Figure \ref{fig:QuadCrushCurves} (thick yellow curve) lead to full consolidation already at a pressure of $\sim$1 MPa. This cannot be supported by laboratory work for any composition, including refractory materials, as shown here and elsewhere, and even pure water ice (see e.g. \cite{MalamudPrialnik-2015}). Thus, the quadratic crush curves typically chosen for comet-like fluffy objects appear highly inaccurate (especially given more energetic collisions). In contrast, the new micro-granular silica crush curve would portray the compression more accurately. By making the crush curve both more exact and also independent of parameters, the simulation outcomes would then become rather more contingent on the choice of other model relations like the cohesion, coefficient of friction, etc. Using the results of this study, the overall uncertainty in model relations may therefore potentially be reduced.

Despite these advantages, the results of our new study are limited to our choice to focus solely on silica, a decision motivated by our objective to isolate the importance of other properties like grain size distribution and texture. Our results are also limited in having fitted our new crush curve only up to the pressures available to us with our new experimental apparatus, which do not exceed 1 GPa by much. Therefore, a lot of future research is still required, with the goal of producing a comprehensive catalog of crush curves encompassing a variety of solar system analog materials, including hydrated refractory minerals, refractory organics, and refractory-ice mixtures. Also required are studies at yet higher pressure, which can reach the limit of full consolidation of silica samples. Additionally, experiments performed at temperatures closer to the solidus temperature of each refractory material are also needed, because then the material is more susceptible to compaction. More realistic solar system materials might consist of extremely wide grain size distribution, as was hinted during the investigation of comet 67P/C-G (e.g., figure 12 in \cite{GuttlerEtAl-2019}), and thus we require new crush curve experiments on samples with similarly ultra-wide grain size distribution. The results of the current study emphasize that the size distribution might be an attribute of fundamental importance.

\section{Conclusions}\label{S:Conclusions}
\begin{itemize}
    \item A crush curve is a mathematical function that describes the change in porosity of a porous material as a function of pressure. It is one of the most important constituent relations that enters into impact simulations and other types of numerical codes relevant to small astrophysical bodies (Section \ref{S:Intro}).

    \item Here we review (Section \ref{S:Past}) as well as supplement (Section \ref{S:Current}) the empirical literature for silica crush curves -- being one of the most frequently used materials in experiments, and highly appropriate to a variety of astrophysical applications.

    \item Our compilation of different crush curves surveys various properties related to the experiments, such as the investigation method (static vs. dynamic compression) grain sizes (and size distributions), textures (spherical vs. angular/irregular) and pressure range. A comparison is drawn between the different resulting curves (Section \ref{S:Comparison}).

    \item We fit the empirical results for micro-granular silica experimentws with an over-arching, power law mathematical function. This new crush curve is compatible with the experiments over a wide range of pressures between a few $10^2$ Pa and up to 1.1 GPa (Figure \ref{fig:CurvesWithWideFit}).

    \item We compare our new crush curve to parametric quadratic crush curves, which are most commonly used in impact simulations of small bodies throughout the literature. We use benchmark impact simulations based on the DART impact, in which we employ an otherwise identical setup, but change only the crush curve (Section \ref{S:Benchmark}).

    \item We find that the benchmark simulations indeed differ for the new crush curve, particularly when the initial porosity of the target increases. The new crush curve effectuates a greater compression in the vicinity of the impact site, and this leads to reduced values of the momentum transfer coefficients (a.k.a. $\beta$; Table \ref{tab:beta}).

    \item We advocate the use of our new crush curve whenever micro-granular silica is an applicable material, because the alternative quadratic crush curve both underestimates and overestimates compression at low and high pressures, respectively. In addition, the new crush curve has a universal rather than case-specific mathematical form, not requiring calibration via ad-hoc parameters (Section \ref{S:Discussion}).

    \item Since the realistic composition and grain properties of asteroid- and comet-like materials are not yet known, we advocate the need to treat crush curves with careful consideration, and conduct further investigations in the meantime, compiling a large catalog of crush curve measurements for various materials, in order to investigate an even wider range of pressures, temperatures and grain size distributions.
    
\end{itemize}

\section{Acknowledgements}
We wish to thank anonymous reviewer for extremely helpful comments that enhanced the quality of the manuscript. We also wish to thank Sabina Raducan for providing us with the latest parameter fits for the LM sand crush curves. UM and HBP acknowledge support from the Israeli Ministry of Science and Technology MOST-space grant and the Minerva Center for Life under extreme conditions. UM, HP, CK and JB acknowledge support by the Niedersächsisches Vorab in the framework of the research cooperation between Israel and Lower Saxony under grant ZN 3630. GM and MT acknowledge support through the German Space Agency (DLR) project 50WM2254A. CK acknowledges support by the European Union under grant agreement NO. 101081937 – Horizon 2022 - Space Science and Exploration Technologies. CMS and CB acknowledge support by the High Performance and Cloud Computing Group at the Zentrum für Datenverarbeitung of the University of Tübingen, the state of Baden-Württemberg through bwHPC and the German Research Foundation (DFG) through grant no INST 37/935-1 FUGG. CB appreciates support by the German Research Foundation (DFG project 285676328).

\section*{Data Availability}

The dataset constructed in this work will be available via a reasonable e-mail request to the lead author.


\newpage
\bibliographystyle{apj}
\bibliography{bibfile} 





\label{lastpage}
\end{document}